\documentclass[reprint, twocolumn, floatfix]{aastex631}

\usepackage[utf8]{inputenc}
\usepackage{amsmath, color, amssymb, latexsym, graphicx, lipsum}
\usepackage{ulem}
\usepackage{color}

\newcommand{\ua}{\textbf{v}_{\text{A}}}

\newcommand{\uh}{\textbf{v}}

\submitjournal{ApJL}
\shorttitle{Jovian Magnetosheath Turbulence}
\shortauthors{Andr\'es et al.}

\begin{document}

\title{Observation of Turbulent Magnetohydrodynamic Cascade in the Jovian Magnetosheath}

\author[0000-0002-1272-2778]{N. Andr\'es}
\affiliation{Departamento de F\'{\i}sica, UBA, Ciudad Universitaria, 1428, Buenos Aires, Argentina}
\affiliation{Instituto de Astronom\'ia y F\'{\i}sica del Espacio, CONICET-UBA, Ciudad Universitaria, 1428, Buenos Aires, Argentina}

\author[0000-0002-6962-0959]{R. Bandyopadhyay}
\affiliation{Department of Astrophysical Sciences, Princeton University, Princeton, NJ 08544, USA}

\author[0000-0001-6160-1158]{D. J. McComas}
\affiliation{Department of Astrophysical Sciences, Princeton University, Princeton, NJ 08544, USA}

\author[0000-0003-2685-9801]{J. R. Szalay}
\affiliation{Department of Astrophysical Sciences, Princeton University, Princeton, NJ 08544, USA}

\author[0000-0003-0696-4380]{F. Allegrini}
\affiliation{Southwest Research Institute, San Antonio, TX, USA}
\affiliation{Department of Physics and Astronomy, University of Texas at San Antonio, San Antonio, TX, USA}

\author[0000-0002-2504-4320]{R. W. Ebert}
\affiliation{Southwest Research Institute, San Antonio, TX, USA}
\affiliation{Department of Physics and Astronomy, University of Texas at San Antonio, San Antonio, TX, USA}

\author[0000-0003-1304-4769]{D. J. Gershman}
\affiliation{Goddard Space Flight Center, Greenbelt, MD, USA}

\author[0000-0001-7478-6462]{J. E. P. Connerney}
\affiliation{Goddard Space Flight Center, Greenbelt, MD, USA}
\affiliation{Space Research Corporation, Annapolis, MD, USA}

\author[0000-0002-9115-0789]{S. J. Bolton}
\affiliation{Southwest Research Institute, San Antonio, TX, USA}


\begin{abstract}
We present the first estimation of the energy cascade rate in Jupiter's magnetosheath (MS). We use in-situ observations from the Jovian Auroral Distributions Experiment (JADE) and the magnetometer investigation (MAG) instruments onboard the Juno spacecraft, in concert with two recent compressible models to investigate the cascade rate in the magnetohydrodynamic (MHD) scales. While a high level of compressible density fluctuations is observed in the Jovian MS, a constant energy flux exists in the MHD inertial range. The compressible isothermal and polytropic energy cascade rates increase in the MHD range when density fluctuations are present. We find that the energy cascade rate in Jupiter's magnetosheath is at least two orders of magnitude (100 times) smaller than the corresponding typical value in the Earth's magnetosheath.
\end{abstract}

\section{Introduction} \label{sec:intro}
Turbulence is a ubiquitous phenomenon observed from quantum to astrophysical scales~{\citep[e.g.,][]{Cl2017,BC2013}}. Briefly, neutral fluid or plasma turbulence  {is mostly} described by a nonlinear  {cascade} of energy, typically from large to small scales. Far from  {the} large and small scales, there exists  {the so-called} inertial range {,} where the energy  {cascade} rate remains constant. In space and astrophysical plasmas, turbulence plays  {an essential} role in  {different} physical processes  {like the} star formation, transport, and plasma heating~\citep[e.g.,][]{R2022,Hu2020,C2017,H2015, Qudsi2020ApJS}.  Magnetospheres of planets such as Earth, Venus, Saturn, and Jupiter are formed by the interaction  {between the magnetized supersonic} solar wind plasma  {and} the  {planet's} intrinsic magnetic field~\citep{R1993}. In particular, these planets have intrinsic magnetic fields of  {enough} strength to stand off the solar wind flow and form a  {planetary} magnetosheath (MS), i.e., a spatial region excluding the solar wind, downstream of the planetary bow shock~\citep{K1995}. A key ingredient of these regions is strong density fluctuations, which in some cases can reach up to 100$\%$ of the mean density~\citep{A2019b,C2017}. These strong density fluctuations are in contrast to the ambient solar wind, which is weakly compressible in general. Jupiter's MS provides a unique laboratory to investigate compressible plasma turbulence. Thanks to the in situ observations provided by the Juno mission in the Jovian MS~\citep{Connerney2017Science, Ranquist2019JGR} and recently derived compressible exact relations, we can estimate the energy cascade rate in the magnetohydrodynamic (MHD) scales around Jupiter for the first time.

{Assuming a stationary state with a finite dissipation rate, a balance between the forcing and the dissipation terms and in the limit of viscosity and resistivity tending to zero}, exact relations (or exact laws) between the energy cascade rate $\varepsilon$ and two-point correlation functions can be obtained \citep{F1995,MY1975}. For fully developed homogeneous, isotropic  {and} incompressible MHD turbulence, the exact relation predicts a linear scaling for a particular longitudinal third-order structure function with the distance $\ell$ between two points \citep{P1998a,P1998b}. The validity of this  {theoretical prediction} has been  {test using direct numerical simulations (DNSs)} \citep[see, e.g.][]{Mi2009}.  {Moreover, this exact} relation has been used to estimate the energy cascade rate in the solar wind and  {different} planetary magnetosheaths \citep{SV2007,Mc2008,B2020,B2020b,A2021}. {For a recent detailed review about scaling laws and exact relations in plasma turbulence see \citet{Ma2023}.}

For compressible MHD turbulence, depending on the closure equation used, different types of exact relations can be derived  {\citep[e.g.][]{B2013, A2017b, S2021, He2021}}. Using the isothermal closure, \citet{A2017b} revisited the work of \citet{B2013} and provided a new derivation using  {a more suitable set of variables for a compressible plasma, i.e., the proton plasma density, the proton plasma velocity field, and the compressible Alfvén speed.} The  {exact} relation reported in \citet{A2017b} showed four types of  {components} that are involved in the nonlinear energy cascade rate:  {the flux and source terms (the same type of terms {involves} in the compressible hydrodynamic (HD) models \citep[see,][]{Ga2011,B2014}) plus two new terms, the hybrid and $\beta$-dependent terms}. Recently, \citet{S2021} proposed a more general method to obtain exact relations for any turbulent isentropic flow. The authors derived an exact  {relation for polytropic MHD turbulence, being this new expression expressed in the same structure as its isothermal counterpart.} The main goal of the present letter is to investigate the rate of transfer of energy $\varepsilon$ at the MHD scales in the compressible Jovian MS plasma using the two recent exact relations derived for fully developed isothermal and polytropic plasma turbulence \citep{A2017b, S2021}.

\section{Exact relations in MHD turbulence}\label{sec:theo}
Using the compressible MHD equations \citep[see,][]{A2017b, S2021},  {and assuming an infinite kinetic and magnetic Reynolds numbers, a stationary state with a match between forcing and dissipation terms, and a finite energy cascade rate, an exact relation can be found as},
\begin{equation}\label{exactlaw}
	-2\varepsilon_C=\frac{1}{2}\boldsymbol\nabla_\ell\cdot\textbf{F}_\text{C}+\mathcal{S},
\end{equation}
where $\varepsilon_C$ represents the compressible energy cascade rate, $\textbf{F}_\text{C}$ is the compressible flux term and $\mathcal{S}$ groups  {the non-flux terms, i.e., the source, hybrid, and $\beta$-dependent terms} \citep[for a detailed derivation of Eq.~\eqref{exactlaw} see,][]{A2017b, S2021}. The  {so-called} flux term is  {related} with the energy flux, and is the  {typical} term  {involve in exact laws of incompressible HD and MHD turbulence \citep{vkh1938,P1998a,P1998b}}.  {These} terms can be written as products of increments of  {the the velocity and magnetic field fields}.  {In particular,} the total compressible flux {\bf F}$_\text{C}$  {in Eq.~\eqref{exactlaw}} is a  {mixture of two different terms}, {\bf F}$_\text{C}$ = {\bf F}$_\text{1C}$ + {\bf F}$_\text{2C}$. The first term is a Yaglom-like term:
\begin{align}\nonumber
	\textbf{F}_\text{1C} \equiv&~ \langle [(\delta(\rho\uh)\cdot\delta\uh+\delta(\rho\ua)\cdot\delta\ua]\delta\uh \\ \label{f1} 
        &- [\delta(\rho\uh)\cdot\delta\ua+\delta\uh\cdot\delta(\rho\ua)]\delta\ua\rangle,
\end{align}
where $\rho$ is the ion mass density, {\bf v} is the ion velocity field and $\ua\equiv\textbf{B}/\sqrt{4\pi\rho}$ is the compressible Alfv\'en velocity. Equation \eqref{f1} is the compressible generalization of the incompressible Yaglom  {component} \citep[see,][]{P1998a,P1998b}.  {Furthermore}, we have a purely compressible flux  {component} {\bf F}$_\text{2C}$, which strongly depends on the closure equation used in the compressible model. As we discussed in the Introduction, in the present  {manuscript} we use two closure equations to complete our compressible models.  {We use an isothermal equation of state $P_\text{iso}\sim\rho$ and a polytropic equation of state $P_\text{pol} \sim \rho^\gamma$, where the polytropic index $\gamma$ is set to $5/3$.}

Using the isothermal closure \citep[see,][]{A2017a}, the compressible flux term is,
\begin{align}\label{f2a}
	\textbf{F}_\text{2C}^\text{iso} \equiv&~ 2\langle \delta\rho\delta e \delta\uh\rangle,
\end{align}
where we have introduced the compressible internal energy for an isothermal plasma $e \equiv c_s^2\ln(\rho/\rho_0)$ (where $\rho_0$  {is the mean mass density and $c_s$ is the constant sound speed}). Using a polytropic closure \citep[see,][]{S2021}, 
\begin{align}\label{f2b}
	\textbf{F}_\text{2C}^\text{poly} \equiv&~ 2\langle \delta\rho\delta u \delta\uh\rangle,
\end{align}
where we have used the internal compressible energy for a polytropic plasma as $u=(c_s^2-c_{s0}^2)/\gamma(\gamma-1)$ {, where $c_s$ is the variable sound speed.}  {According to Eqs.~\eqref{f2a} and \eqref{f2b}, there are new components in the total energy cascade rate due only to the presence of proton density fluctuations in the plasma \citep[see,][]{B2013,A2017b,S2021}}. 

 {In the previous equations, the prime notation indicates field evaluation at $\textbf{x}'=\textbf{x}+\boldsymbol\ell$, where $\boldsymbol\ell$ is the increment vector. The angular bracket $\langle\cdot\rangle$ indicates a typical ensemble average. Also, assuming ergodicity and spatial homogeneity the results of averaging over a large number of realizations can be obtained equally well by averaging over a large region of space (or time assuming the Taylor's hypothesis) for one realization \citep{F1995, Taylor1938PRSLA}.} 

The  {other energy cascade components} included in $\mathcal{S}$ in Eq.~\eqref{exactlaw} are a combination of two-point correlation functions proportional to the divergence of the  {compressible Alfv\'en and velocity fields}.  {These terms may act as a source or a sink for the energy cascade rate in the inertial range \citep{Ga2011,A2017b}. Also, it is worth mentioning that the terms in $\mathcal{S}$ can be computed only using multi-spacecraft observations since, as we discussed previously, those include local vector divergences \citep[see,][]{A2018}.}  {In the present manuscript, we are dealing with single-spacecraft observations, then we cannot estimate the non-flux terms. However, previous observational works using MMS measurements in the Earth's MS shown that the non-flux components were indeed negligible with respect to the flux terms in the inertial range \citep{A2019b}. Moreover, compressible MHD simulations for fully developed turbulence also supported this result \citep{A2018,F2020}.} Therefore, in the present work,  {to estimate the compressible energy cascade rate \eqref{exactlaw} in the inertial range, we consider only Eqs.~\eqref{f1}, \eqref{f2a} and \eqref{f2b}, i.e., the compressible MHD flux terms.}

 {Assuming statistical full isotropy, it is possible to integrate Eq.~\eqref{exactlaw} to obtain a scalar relation valid for isotropic turbulence as,}
\begin{align}\nonumber
\varepsilon^\text{iso}_\text{C}=&~\varepsilon_\text{1C}+\varepsilon^\text{iso}_\text{2C} \\ \label{iso} =& \langle\text{F}_\text{1C}/(-4/3V\tau)\rangle+\langle\text{F}^\text{iso}_\text{2C}/(-4/3V\tau)\rangle, \\ \nonumber 
	\varepsilon^\text{poly}_\text{C}=&~\varepsilon_\text{1C}+\varepsilon^\text{poly}_{2C} \\ =& \label{pol} \langle\text{F}_\text{1C}/(-4/3V\tau)\rangle+\langle\text{F}^\text{poly}_\text{2C}/(-4/3V\tau)\rangle,
\end{align}
where the flux terms have been projected into the mean solar wind flow velocity field $V\equiv\langle|{\bf v}|\rangle$ and we have assumed the Taylor's hypothesis  {(i.e., $\ell\equiv V\tau$)}.  {Finally,} the incompressible limit is easily recovered for $\rho\rightarrow\rho_0$,
\begin{align}\nonumber
\varepsilon_\text{I} &= \langle\text{F}_\text{I}/(-4/3V\tau)\rangle \\ \label{inc} 
&= \rho_0\big\langle \{[(\delta\uh)^2+(\delta{\bf v}_\text{A})^2]\delta\uh - 2(\delta\uh\cdot\delta{\bf v}_\text{A})\delta{\bf v}_\text{A}\}/(-4/3V\tau)\big\rangle.
\end{align}
 {where $\varepsilon_\text{I}$ is the incompressible transfer of energy that cascades through the MHD inertial range \citep[see,][]{P1998a,P1998b}.}

\section{Results}\label{sec:res}

\subsection{Juno observations}

The Juno orbits \citep[see,][]{E2017, Ho2017, G2017, Ranquist2019JGR} provide us a first opportunity to estimate the incompressible and compressible energy cascade rates in Jupiter's dawn-side MS through the use of magnetic and plasma in situ observations \citep[see,][]{Co2017, Co2017b, Mc2017}. The Juno magnetic field investigation (MAG) consists of two independent magnetometer sensor suites, each consisting of a tri-axial fluxgate magnetometer sensor and a pair of co-located imaging sensors mounted on an ultra-stable optical bench. The Jovian Auroral Distributions Experiment (JADE) is an instrumental suite that includes electron sensors (JADE-E) and a single ion sensor (JADE-I), which we use in the present work \citep{Mc2017}. JADE-I measures ions from $\sim$ 13 eV/q to $\sim$ 46 keV/q. The observations analyzed in this study are from the low-rate mode, recorded every 30 s in the MS.

\begin{figure}[ht]
\begin{center}
\includegraphics[width=\linewidth]{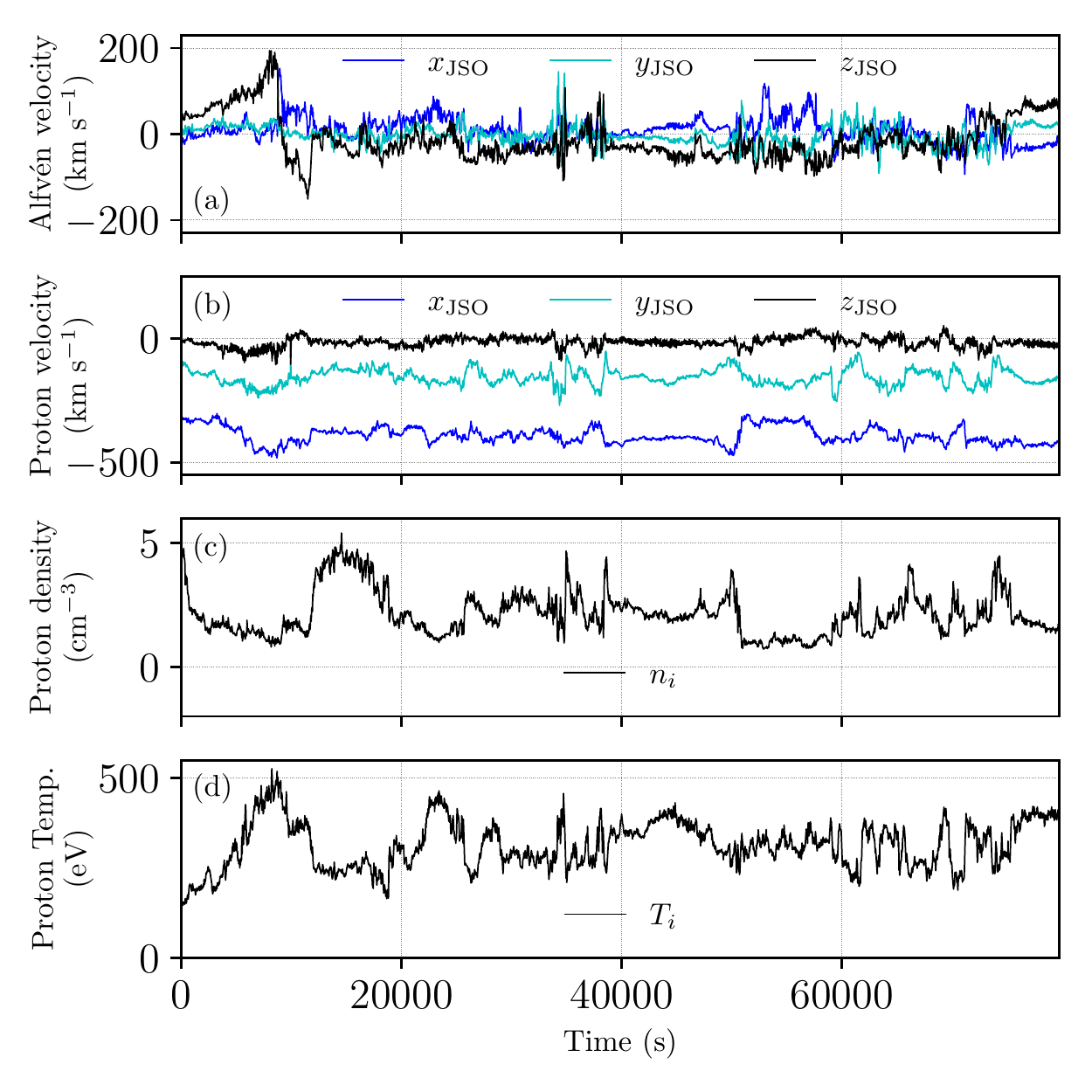}
\end{center}
\caption{Compressible Alfv\'en velocity (in Jupiter-Sun-Orbit [JSO] coordinates), proton velocity (JSO), proton number density fluctuations and proton temperature measurements in Jupiter's magnetosheath.}
\label{fig:0}
\end{figure}

To investigate the compressible MHD turbulence in the Jovian MS, we used Juno's observations analyzed in detail recently by \citet{Ba2021}. In particular, we consider  {observations} in the dawn-side of the  {Jovian MS} obtained by Juno during a  {MS} crossing on February 20, 2017 from 01:00:23 to 23:09:54 UTC.  {This particular event was at a distance of $\sim$ 105 $R_J$, where $R_J$ is the Jovian Radius equal to 71492 km.} Figure \ref{fig:0} shows (a) the compressible Alfvén velocity in the Jupiter-Sun-Orbit (JSO) coordinate system (where $x_\text{JSO}$ is the direction from the center of Jupiter  {to the center of the Sun}, $y_\text{JSO}$ is the direction opposite to Jupiter's  {orbital} motion around the Sun, and the $z_\text{JSO}$ completes the right-handed coordinate system, see \citet{Ba2017}), (b) the proton velocity field components (JSO), (c) the number density and (d) the temperature as a function of time, respectively.  {We focus on this particular interval as i) it is a very long time interval relative to other different JADE MS samples and ii) there is a continuous and high-resolution number of observations available.} This  {two reasons} allows us to investigate a wide range  {of spatial and temporal} scales. Furthermore, there is no clear large-scale discontinuity, which allows us to study the energy cascade rates, derived from homogeneous MHD theories.

\subsection{Turbulence Spectra} \label{sec:spectra}

To explore the nature of  {the MHD turbulent inertial range}, we begin by computing the power spectra of  {the} magnetic and ion velocity  {fields} for the  {MS} interval.  {In the presence of a mean velocity flow (relative to the spacecraft) sufficiently faster than the Alfv\'en wave speed, one can assume the Taylor's frozen-in hypothesis~\citep{Taylor1938PRSLA}. Using this hypothesis, we are able to interpret the temporal fluctuations at time scales $\tau$ as spatial fluctuations at spatial scales $\ell$ using $\ell = V \tau$, where $V$ is the flow speed of the plasma. For the previously described Juno event,} the Alfv\'en speed,  {computed} from the mean magnetic field and  {mean} mass density, is about 26 times smaller than the mean flow speed and the ratio of fluctuation to mean flow is also rather small: $v_{\mathrm{rms}}/V\sim0.1$ (where, $v_\text{rms} \equiv \sqrt{\langle|{\bf v}-\langle{\bf v}\rangle|^2\rangle}$). Therefore, we can assume that Taylor's frozen-in approximation is valid for this case.

\begin{figure}
	\begin{center}
		\includegraphics[width=0.85\linewidth]{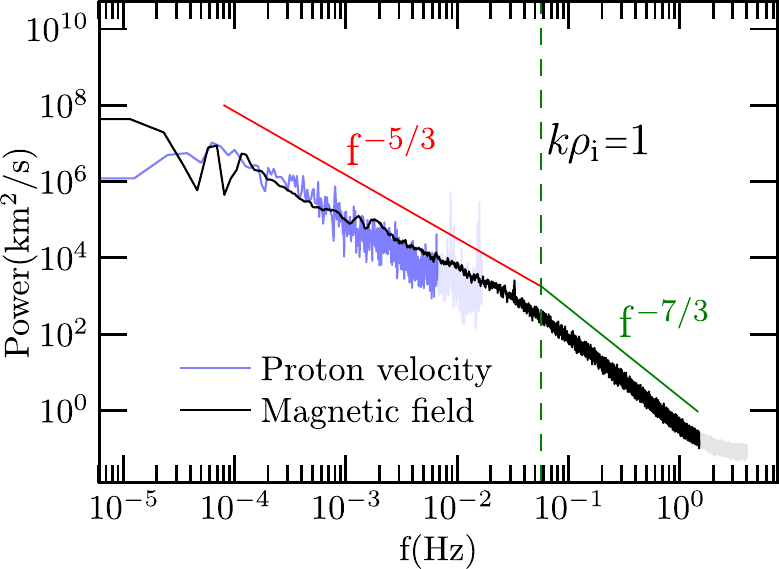}
		\caption{ {Spectra for the magnetic field (solid, black) and proton velocity field (faded, blue) as a function of the frequency. A Kolmogorov scaling $f^{-5/3}$} is shown for reference.  {In addition,} the dashed, green vertical line represents the ion gyro-radius scale $k \rho_{\mathrm{i}} = 1$ with the wave vector $k \simeq (2\pi f)/|\langle \mathbf{V} \rangle|$. In both cases, the high-frequency part of the spectra, where instrumental artifacts become prominent, have  been further lightly shaded.}
		\label{fig:spec}
	\end{center}
\end{figure}

Figure~\ref{fig:spec} shows the  {spectrum for the magnetic field} (faded, blue) computed from the MAG  {observations} and  {the spectrum for the ion velocity field} (solid, black) from JADE  {observations}. The magnetic field data have been converted to Alfv\'en units. The frequency has been converted to wavenumber assuming Taylor's hypothesis. Both magnetic and kinetic spectra exhibit a clear inertial range with a powerlaw close to Kolmogorov $-5/3$ scaling.  {The magnetic-field spectrum appears to have a shallower slope than that of the velocity field in the inertial range. From this single case study, it is not possible to conclude whether this is the typical case in Jupiter’s magnetosheath. A larger survey with more intervals will clarify this.} {In the case of the magnetic field spectrum, it shows a steepening $\sim -7/3$ slope beginning near the ion-gyro radius $\rho_{\mathrm{i}}$, indicating the presence of dissipative processes {\citep{BC2013}}.} The flattening of the magnetic field spectrum near 1.5 Hz is likely noise-dominated, as the signal-to-noise ratio becomes smaller than 3 here (see Appendix \ref{app}). The velocity spectrum does not show the kinetic-range steepening due to relatively low temporal cadence.

The broadband Kolmogorov scaling of the magnetic and velocity field indicates a  {a classical Kolmogorov fully developed turbulence with a scale invariant flux of energy through the MHD inertial range}~\citep{F1995, Biskamp2003Book, Huang2014ApJL}. In the next section, we provide direct evidence of this cascade and a quantitative estimate of the strength of the energy transfer rate via third-order statistics.

\subsection{The energy cascade rates}
\begin{figure}[ht]
\begin{center}
\includegraphics[width=\linewidth]{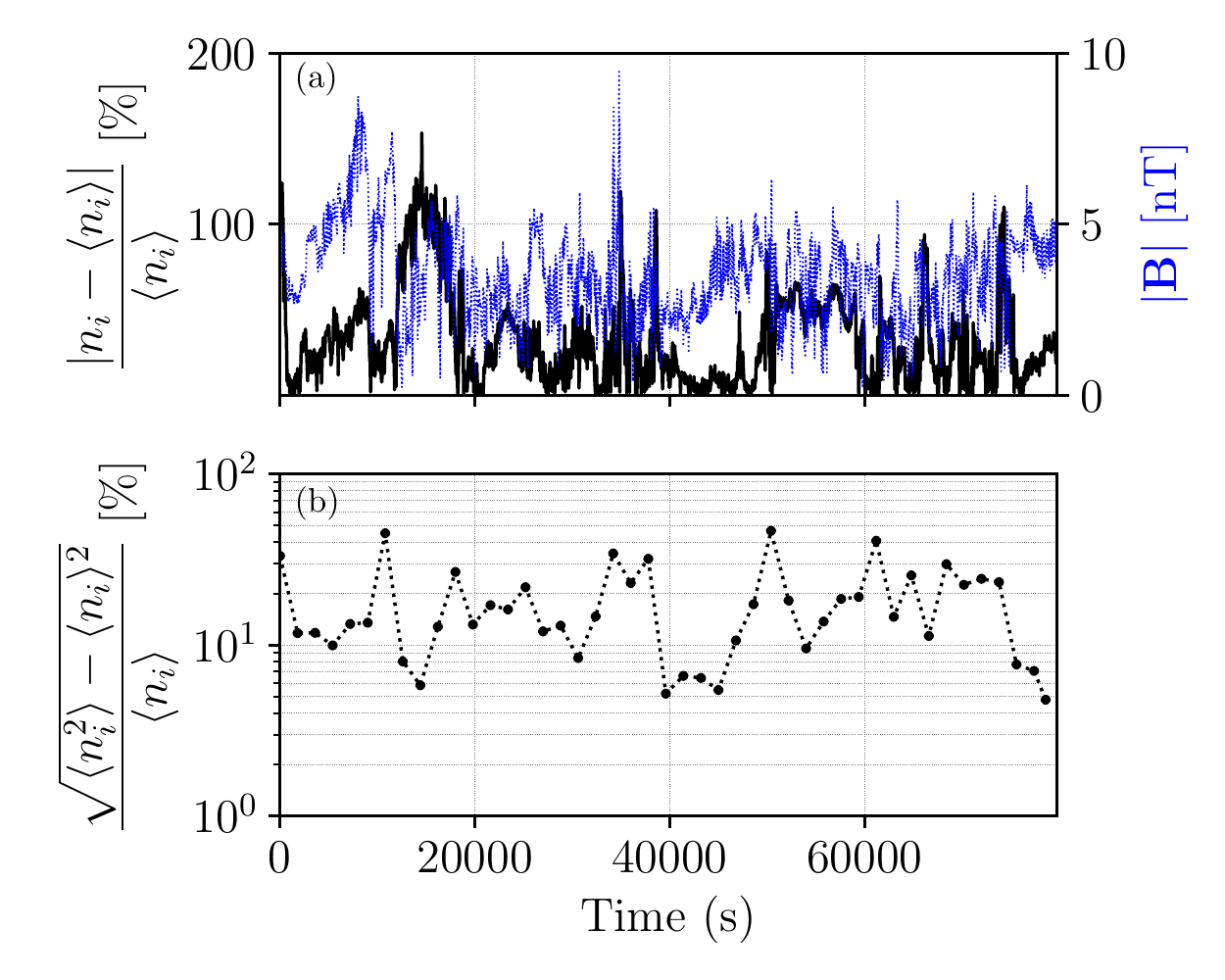}
\end{center}
\caption{(a) Number density fluctuations and compressibility ratio $\sqrt{\langle n_i^2 \rangle-\langle n_i \rangle^2}/\langle n_i \rangle$ compute every 30 minutes as a function of time.}
\label{fig:1}
\end{figure}

Figure \ref{fig:1} (a) shows the number density fluctuations $|n_i - \langle n_i \rangle|/\langle n_i \rangle$ and the absolute magnetic value $|{\bf B}|$ as a function of time.
In particular, we observe i) ranges where the density fluctuations are of the order of the mean number density; ii) $\langle n_i \rangle$ and $|{\bf B}|$ present strong fluctuation levels. Both observational results are typical for magnetosonic like events \citep[see,][]{H2017b,Hu2017,A2019b}. In addition, Figure \ref{fig:1} (b) shows the compressibility ratio (defined as $\sqrt{\langle n_i^2 \rangle-\langle n_i \rangle^2}/\langle n_i \rangle$) computed every 30 minutes as a function of time. Like other planetary MSs \citep[see,][]{H2015,Hu2017}, Figure \ref{fig:1} confirms the presence of large compressibility intervals in Jupiter's MS. In particular, this result makes the Jovian MS an excellent candidate to investigate the compressible energy cascade rate.

For the selected time interval in Figure \ref{fig:0}, we compute $\varepsilon$ from the isothermal, polytropic and incompressible models using Eqs.~\eqref{iso}, \eqref{pol} and \eqref{inc}{, respectively}. For this, we constructed temporal correlation functions of the different turbulent variables in those equations, varying the time lag $\tau$ from 30 s to 65280 s {(or from 0.5 h to 1088 h)}, allowing us to be well inside the MHD large scales. 
\begin{figure}[ht]
\begin{center}
\includegraphics[width=\linewidth]{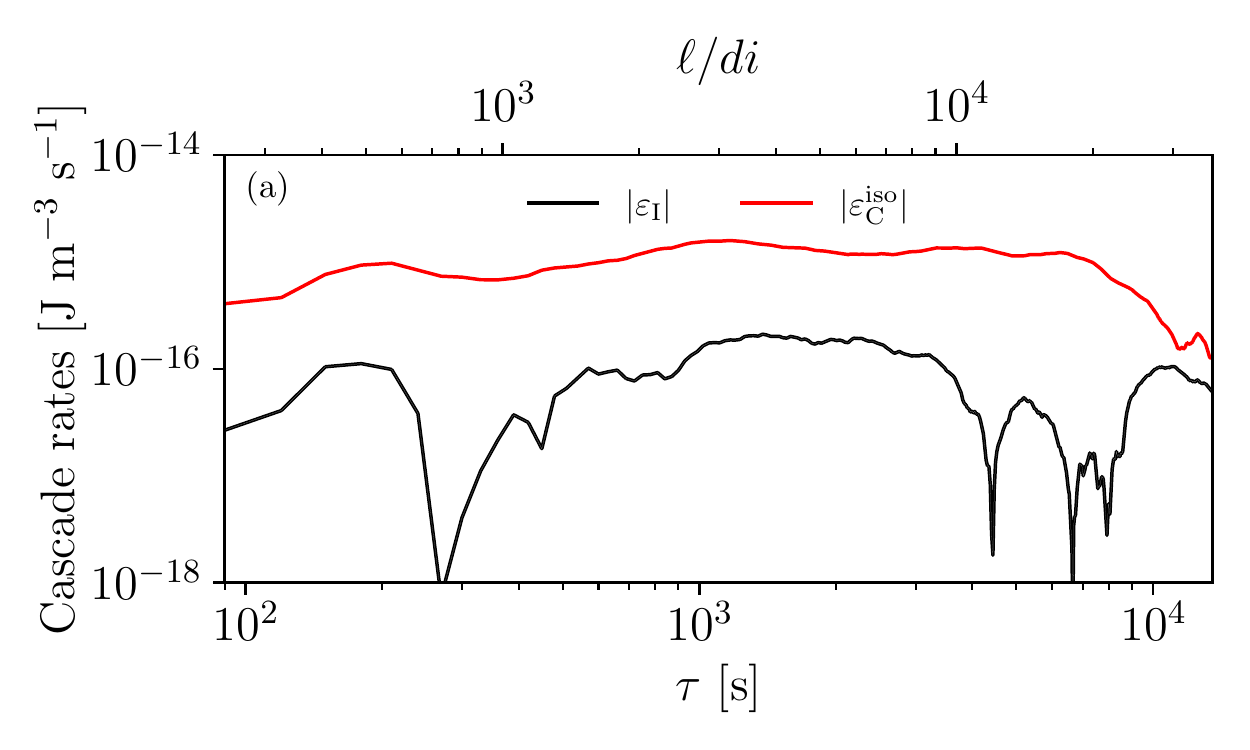}\\
\includegraphics[width=\linewidth]{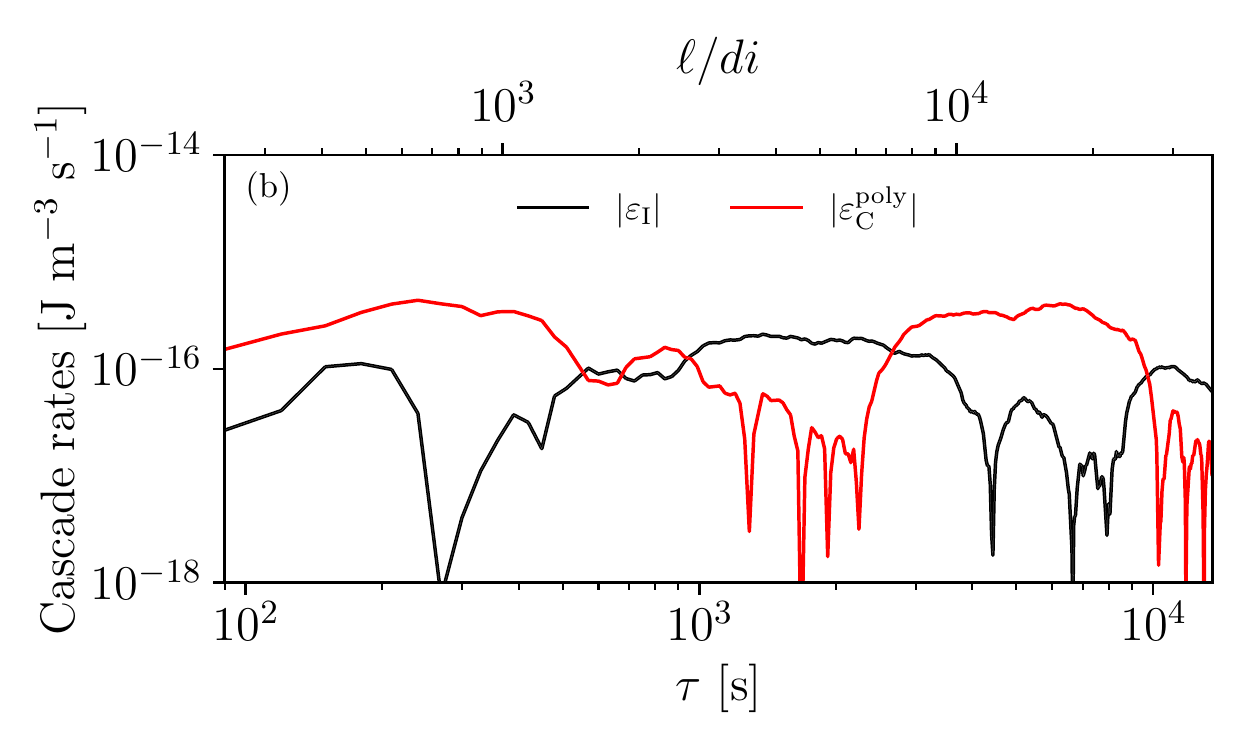}\\
\end{center}
\caption{Energy cascade rate in the Jovian magnetosheath using the isothermal (red, top panel) and the polytropic closure (red, bottom panel) as a function of time and spatial lags (using the Taylor hypothesis), respectively. The incompressible cascade rate is shown using black solid lines in both panels.}
\label{fig:2}
\end{figure}
Figure \ref{fig:2} shows the absolute value of the incompressible and the compressible energy cascade rates in the Jovian MS using (a) the isothermal and (b) the polytropic closure as a function of time and spatial lags (assuming Taylor's hypothesis), respectively. Interestingly, {for the isothermal case we observe a constant value of the absolute value of the compressible cascade rate  $|\varepsilon_\text{C}^\text{iso}|$ for more than a decade on scales. Moreover, we noted a clear increase of $|\varepsilon_\text{C}^\text{iso}|$ (at least one order of magnitude) with respect to the incompressible cascade rate $|\varepsilon_\text{I}|$ when density fluctuations are taken into account.}

In addition, we investigate the signed components of the compressible cascade rate through the use of Eqs.~\eqref{f1}, \eqref{f2a} and \eqref{f2b}. 
\begin{figure}[ht]
\begin{center}
\includegraphics[width=\linewidth]{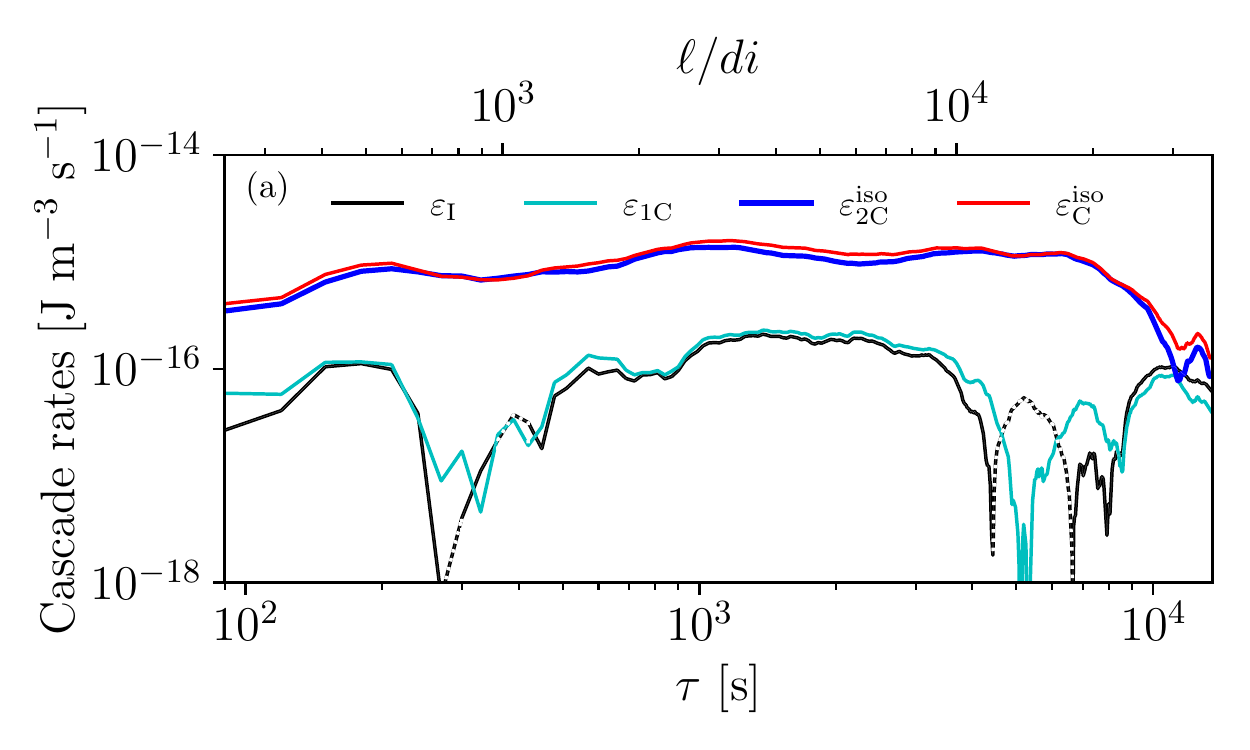} \\
\includegraphics[width=\linewidth]{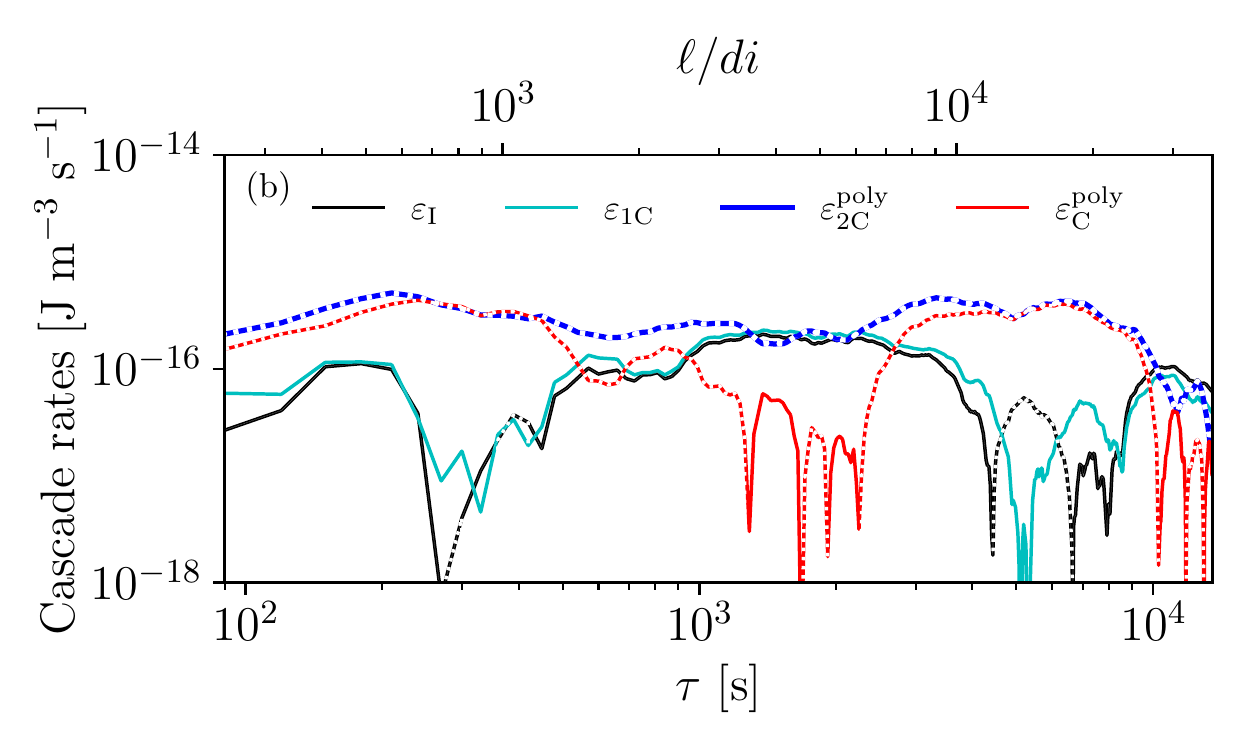}
\end{center}
\caption{Energy cascade rate components for the isothermal (top) and the polytropic closure (bottom) as a function of time and spatial lags (using the Taylor hypothesis), respectively. {In addition, in both panels, the incompressible cascade rate is reported.} Dashed lines correspond to negative values and solid lines to positive values. }
\label{fig:3}
\end{figure}
Figure \ref{fig:3} shows the {signed} compressible components for (a) the isothermal and (b) polytropic models, respectively. {For the sake of comparison, in both panels (a) and (b) the signed incompressible cascade component is in black line.} While for isothermal model, the energy cascade is due to the compressible component $\varepsilon^\text{iso}_\text{2C}$ (direct cascade), for the polytropic model, there is a competition between both flux components $\varepsilon_\text{1C}$ and $\varepsilon_\text{2C}$ though all the inertial range. {We note that $\varepsilon_\text{1C}$ and $\varepsilon^\text{poly}_\text{2C}$ have a few constant scales (less than a decade), which could be due to some specific scale features or simply to insufficiently statistical convergence of the third order component \citep[see,][]{Ma2023}. This behavior has two consequences, i) break the constant polytropic energy cascade rate and ii) the total polytropic cascade rate is of the order of the incompressible cascade rate. Moreover, it is worth mentioning that the incompressible cascade is quite similar to the Yaglom-like generalization $\varepsilon_\text{1C}$, which implies that the large density fluctuation present in the Jovian MS are not affecting the $\varepsilon_\text{1C}$ component.} In both cases, we conclude that the purely compressible cascade rate $\varepsilon_\text{2C}$ is a crucial ingredient to estimate the total turbulent cascade rate in the Jovian MS. 

\section{Discussion and Conclusions}\label{sec:dis}

To the best of our knowledge, we estimate for the first time the incompressible and compressible energy cascade rate in the Jovian MS. Using Juno magnetic and plasma moment observations and exact relations, we observe that deep in the MHD scales the compressible rate is increased when plasma density fluctuations are included in the theoretical description. Previous observational results in different planetary MS \citep{H2017b,Ba2018,A2019b,B2019,B2020b} and in the solar wind \citep{B2016c,H2017a,A2021,S2021} have performed similar analyses both in the MHD scales and in the ion scales. Comparing Jupiter's and Earth's MS, we observe that both  incompressible and compressible energy cascade rates in the Terrestrial MS \citep{H2017b,A2019b} is at least two orders of magnitude larger those in the Jovian MS. 
\begin{table}
	\caption{Typical values of {absolute value of} {the incompressible ($ \varepsilon_{I} $) and compressible ($ \varepsilon_{C} $) energy cascade rates estimate in Jupiter's magnetosheath, Earth's magnetosheath and the solar wind at 1 au.}}
	\label{tab:eps}
	\begin{center}
		\begin{tabular}{c c c}
			\hline \hline
			System 
			& $ |\varepsilon_{I}| $ & $ |\varepsilon_{C}| $   \\
			& \colhead{($\rm W\,m^{-3}$)} & \colhead{($\rm W\,m^{-3}$)}   \\
			\hline
                Jupiter's  {MS} & $\sim 10^{-16}$ & $\sim 10^{-15}$  \\
			Earth's  {MS} & $\sim 10^{-14} - 10^{-13}$ & $\sim 10^{-12} - 10^{-11}$  \\
		    Solar Wind at 1 au & $\sim 10^{-17}$ & $\sim 10^{-17} - 10^{-16}$  \\
			\hline
		\end{tabular}
	\end{center}
\end{table}
We summarize the comparison in Table~\ref{tab:eps}.

For both compressible closures (isothermal \citep{A2017b} and polytropic \citep{S2021}) we observe that the leading components in the cascade estimation \eqref{exactlaw} are the compressible flux terms \eqref{f2a} and \eqref{f2b}, respectively. Previous numerical \citep{A2018b,F2020} and observational \citep{A2021,S2021} results have shown that the compressible cascade rate $\varepsilon_2$ increase with respect to the Yaglom-like cascade $\varepsilon_1$ when compressibility increases in the plasma. These results suggest that in the case that compressible fluctuations are large enough, the compressible flux term could determine the total energy cascade rate. However, in these previous studies the plasma, compressibility was less than 15-20 $\%$ (i.e., for small sonic Mach number runs and in the pristine solar wind) \citep[see,][]{A2017a,A2018b,A2021}. In this paper, for some temporal ranges, the compressibility is larger than 50 $\%$. Therefore, our observational results report here confirms that the leading role in the estimation of the energy cascade rate is due to the compressible terms. It is worth mentioning that for the polytropic relation, the cascade rate is a competition between both components (direct and inverse), in contrast to the isothermal relation, where the compressible term is dominant over the Yaglom-like term in all the MHD scales (both direct components). To conclude if this is a statistically relevant result, more observations with compressible events are required.

The energy cascade rate, estimated using the various third-order exact laws, provide an estimation of the heating rate due to dissipation of the turbulent fluctuations. These results will help in providing quantitative inputs for planetary magnetosheath models.  Solar wind turbulence has been extensively investigated with the help of a large  number of single- and multi-spacecraft missions. However,  a detailed analysis of the properties of the turbulent process in planetary magnetosheaths have only began recently. Characterizing properties of turbulence across diverse plasma environments is critical to develop a unified description of the turbulence physics and phenomenology that is applicable to a variety of different systems. Our results make progress towards this goal.

\section{Acknowledgements}
The authors would like to thank all the Juno team members that made these observations possible. N.A. acknowledges financial support from the following grants: PICT 2018 1095 and UBACyT 20020190200035BA. The research at Princeton University was supported by NASA through Contract NNM06AA75C with the Southwest Research Institute. The JADE-I ion species data are part of the {JNO-J/SW-JAD-3-CALIBRATED-V1.0} data set (ion species data, V02 files) and were obtained from the Planetary Data System (PDS) at \url{https://pds-ppi.igpp.ucla.edu/search/view/?id=pds://PPI/JNO-J_SW-JAD-3-CALIBRATED-V1.0}. The MAG data were obtained from the PDS at \url{https://pds-ppi.igpp.ucla.edu/search/view/?id=pds://PPI/JNO-J-3-FGM-CAL-V1.0}.




\newpage
\appendix

\section{Magnetic-Field Noise}\label{app}

Figure~\ref{fig:spec} (in the main text) shows a flattening at the end of the magnetic field spectrum, which could be due to the fact that the fluctuations are close to noise. To assess the impact of the noise level in the magnetic field data, especially in the high frequency range, we use the magnetic field spectrum in the pristine solar wind. The level of fluctuations being relatively small, the solar wind spectrum is usually a good proxy for the noise level. 

\begin{figure}[hb]
	\begin{center}
		\includegraphics[scale=0.8]{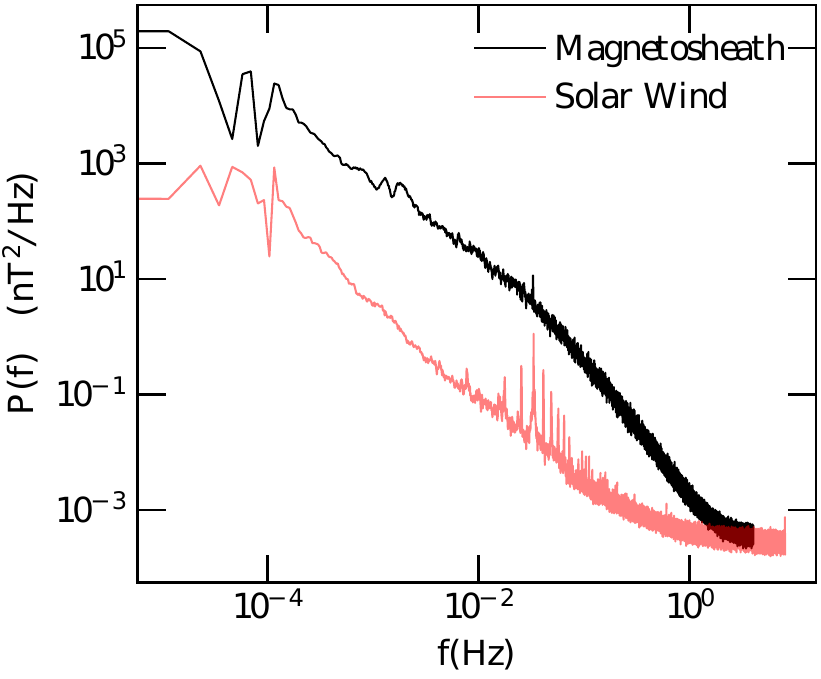}
		\caption{Power spectral density of magnetic-field measurement. The faded red curve is a spectrum measured in the solar wind, considered here to represent the upper bound of the sensitivity floor of the instrument.}
		\label{fig:noise}
	\end{center}
\end{figure}

In Figure~\ref{fig:noise}, we compare the MS spectrum with respect to the one in the solar wind measured by the MAG instrument onboard Juno. The solar wind sample is taken for the interval 2016-05-25T00:00:00.918 to 2016-05-26T00:00:00.749 UTC, when the spacecraft was at $\sim 11356$ RJ from Jupiter towards the Sun, well outside the bow shock perturbations. A few spin tones and harmonics in the solar wind magnetic data can be seen due to instrumental artifacts. The ratio of the two spectral amplitudes becomes lower than 3 above frequency $\sim 1.5$ Hz.

\clearpage


\begin{thebibliography}{}
\expandafter\ifx\csname natexlab\endcsname\relax\def\natexlab#1{#1}\fi
\providecommand{\url}[1]{\href{#1}{#1}}
\providecommand{\dodoi}[1]{doi:~\href{http://doi.org/#1}{\nolinkurl{#1}}}
\providecommand{\doeprint}[1]{\href{http://ascl.net/#1}{\nolinkurl{http://ascl.net/#1}}}
\providecommand{\doarXiv}[1]{\href{https://arxiv.org/abs/#1}{\nolinkurl{https://arxiv.org/abs/#1}}}

\bibitem[{Andr{\'e}s {et~al.}(2017)Andr{\'e}s, Clark~di Leoni, Mininni,
  Dmitruk, Sahraoui, \& Matthaeus}]{A2017a}
Andr{\'e}s, N., Clark~di Leoni, P., Mininni, P.~D., {et~al.} 2017, Physics of
  Plasmas, 24, 102314

\bibitem[{Andr{\'e}s {et~al.}(2018)Andr{\'e}s, Galtier, \& Sahraoui}]{A2018}
Andr{\'e}s, N., Galtier, S., \& Sahraoui, F. 2018, Physical Review E, 97,
  013204

\bibitem[{Andr{\'e}s \& Sahraoui(2017)}]{A2017b}
Andr{\'e}s, N., \& Sahraoui, F. 2017, Physical Review E, 96, 053205

\bibitem[{Andr\'es {et~al.}(2019)Andr\'es, Sahraoui, Galtier, Hadid, Ferrand,
  \& Huang}]{A2019b}
Andr\'es, N., Sahraoui, F., Galtier, S., {et~al.} 2019, Phys. Rev. Lett., 123,
  245101, \dodoi{10.1103/PhysRevLett.123.245101}

\bibitem[{Andr{\'e}s {et~al.}(2021)Andr{\'e}s, Sahraoui, Hadid, Huang,
  Romanelli, Galtier, Dibraccio, \& Halekas}]{A2021}
Andr{\'e}s, N., Sahraoui, F., Hadid, L., {et~al.} 2021, The Astrophysical
  Journal, 919, 19

\bibitem[{Andrés {et~al.}(2018)Andrés, Sahraoui, Galtier, Hadid, Dmitruk, \&
  Mininni}]{A2018b}
Andrés, N., Sahraoui, F., Galtier, S., {et~al.} 2018, Journal of Plasma
  Physics, 84, 905840404, \dodoi{10.1017/S0022377818000788}

\bibitem[{Bagenal {et~al.}(2017)Bagenal, Adriani, Allegrini, Bolton, Bonfond,
  Bunce, Connerney, Cowley, Ebert, Gladstone, {et~al.}}]{Ba2017}
Bagenal, F., Adriani, A., Allegrini, F., {et~al.} 2017, Space Science Reviews,
  213, 219

\bibitem[{Bale {et~al.}(2019)Bale, Badman, Bonnell, Bowen, Burgess, Case,
  Cattell, Chandran, Chaston, Chen, {et~al.}}]{B2019}
Bale, S., Badman, S., Bonnell, J., {et~al.} 2019, Nature, 576, 237

\bibitem[{Bandyopadhyay {et~al.}(2021)Bandyopadhyay, McComas, Szalay,
  Allegrini, Bolton, Ebert, Wilson, \& Gershman}]{Ba2021}
Bandyopadhyay, R., McComas, D., Szalay, J., {et~al.} 2021, Geophysical Research
  Letters, 48, e2021GL095006

\bibitem[{Bandyopadhyay {et~al.}(2018)Bandyopadhyay, Chasapis, Chhiber,
  Parashar, Matthaeus, Shay, Maruca, Burch, Moore, Pollock, {et~al.}}]{Ba2018}
Bandyopadhyay, R., Chasapis, A., Chhiber, R., {et~al.} 2018, The Astrophysical
  Journal, 866, 106

\bibitem[{Bandyopadhyay {et~al.}(2020{\natexlab{a}})Bandyopadhyay, Goldstein,
  Maruca, Matthaeus, Parashar, Ruffolo, Chhiber, Usmanov, Chasapis, Qudsi,
  {et~al.}}]{B2020}
Bandyopadhyay, R., Goldstein, M., Maruca, B., {et~al.} 2020{\natexlab{a}}, The
  Astrophysical Journal Supplement Series, 246, 48

\bibitem[{Bandyopadhyay {et~al.}(2020{\natexlab{b}})Bandyopadhyay,
  Sorriso-Valvo, Chasapis, Hellinger, Matthaeus, Verdini, Landi, Franci,
  Matteini, Giles, Gershman, Moore, Pollock, Russell, Strangeway, Torbert, \&
  Burch}]{B2020b}
Bandyopadhyay, R., Sorriso-Valvo, L., Chasapis, A., {et~al.}
  2020{\natexlab{b}}, Phys. Rev. Lett., 124, 225101,
  \dodoi{10.1103/PhysRevLett.124.225101}

\bibitem[{Banerjee \& Galtier(2013)}]{B2013}
Banerjee, S., \& Galtier, S. 2013, Physical Review E, 87, 013019

\bibitem[{Banerjee \& Galtier(2014)}]{B2014}
---. 2014, Journal of Fluid Mechanics, 742, 230

\bibitem[{Banerjee {et~al.}(2016)Banerjee, Hadid, Sahraoui, \&
  Galtier}]{B2016c}
Banerjee, S., Hadid, L.~Z., Sahraoui, F., \& Galtier, S. 2016, The
  Astrophysical Journal Letters, 829, L27

\bibitem[{Biskamp(2003)}]{Biskamp2003Book}
Biskamp, D. 2003, Magnetohydrodynamic Turbulence (Cambridge, UK: CUP)

\bibitem[{Bruno \& Carbone(2013)}]{BC2013}
Bruno, R., \& Carbone, V. 2013, Living Reviews in Solar Physics, 10, 1

\bibitem[{Chen \& Boldyrev(2017)}]{C2017}
Chen, C.~H., \& Boldyrev, S. 2017, The Astrophysical Journal, 842, 122

\bibitem[{Clark~di Leoni {et~al.}(2017)Clark~di Leoni, Mininni, \&
  Brachet}]{Cl2017}
Clark~di Leoni, P., Mininni, P.~D., \& Brachet, M.~E. 2017, Phys. Rev. A, 95,
  053636, \dodoi{10.1103/PhysRevA.95.053636}

\bibitem[{Connerney {et~al.}(2017{\natexlab{a}})Connerney, Benn, Bjarno,
  Denver, Espley, Jorgensen, Jorgensen, Lawton, Malinnikova, Merayo,
  {et~al.}}]{Co2017}
Connerney, J., Benn, M., Bjarno, J., {et~al.} 2017{\natexlab{a}}, Space Science
  Reviews, 213, 39

\bibitem[{Connerney {et~al.}(2017{\natexlab{b}})Connerney, Adriani, Allegrini,
  Bagenal, Bolton, Bonfond, Cowley, Gerard, Gladstone, Grodent,
  {et~al.}}]{Co2017b}
Connerney, J.~E., Adriani, A., Allegrini, F., {et~al.} 2017{\natexlab{b}},
  Science, 356, 826

\bibitem[{{Connerney} {et~al.}(2017){Connerney}, {Adriani}, {Allegrini},
  {Bagenal}, {Bolton}, {Bonfond}, {Cowley}, {Gerard}, {Gladstone}, {Grodent},
  {Hospodarsky}, {Jorgensen}, {Kurth}, {Levin}, {Mauk}, {McComas}, {Mura},
  {Paranicas}, {Smith}, {Thorne}, {Valek}, \& {Waite}}]{Connerney2017Science}
{Connerney}, J.~E.~P., {Adriani}, A., {Allegrini}, F., {et~al.} 2017, Science,
  356, 826, \dodoi{10.1126/science.aam5928}

\bibitem[{Ebert {et~al.}(2017)Ebert, Allegrini, Bagenal, Bolton, Connerney,
  Clark, DiBraccio, Gershman, Kurth, Levin, {et~al.}}]{E2017}
Ebert, R.~W., Allegrini, F., Bagenal, F., {et~al.} 2017, Geophysical Research
  Letters, 44, 4401

\bibitem[{Ferrand {et~al.}(2020)Ferrand, Galtier, Sahraoui, \&
  Federrath}]{F2020}
Ferrand, R., Galtier, S., Sahraoui, F., \& Federrath, C. 2020, The
  Astrophysical Journal, 904, 160, \dodoi{10.3847/1538-4357/abb76e}

\bibitem[{Frisch(1995)}]{F1995}
Frisch, U. 1995, Turbulence: The Legacy of A. N. Kolmogorov (Cambridge
  University Press.)

\bibitem[{Galtier \& Banerjee(2011)}]{Ga2011}
Galtier, S., \& Banerjee, S. 2011, Physical review letters, 107, 134501

\bibitem[{Gershman {et~al.}(2017)Gershman, DiBraccio, Connerney, Hospodarsky,
  Kurth, Ebert, Szalay, Wilson, Allegrini, Valek, {et~al.}}]{G2017}
Gershman, D.~J., DiBraccio, G.~A., Connerney, J.~E., {et~al.} 2017, Geophysical
  Research Letters, 44, 7559

\bibitem[{Hadid {et~al.}(2017)Hadid, Sahraoui, \& Galtier}]{H2017a}
Hadid, L., Sahraoui, F., \& Galtier, S. 2017, The Astrophysical Journal, 838, 9

\bibitem[{Hadid {et~al.}(2018)Hadid, Sahraoui, Galtier, \& Huang}]{H2017b}
Hadid, L., Sahraoui, F., Galtier, S., \& Huang, S. 2018, Phys. Rev. Lett., 120,
  055102

\bibitem[{Hadid {et~al.}(2015)Hadid, Sahraoui, Kiyani, Retino, Modolo, Canu,
  Masters, \& Dougherty}]{H2015}
Hadid, L., Sahraoui, F., Kiyani, K.~H., {et~al.} 2015, The Astrophysical
  journal letters, 813, L29

\bibitem[{Hellinger {et~al.}(2021)Hellinger, Papini, Verdini, Landi, Franci,
  Matteini, \& Montagud-Camps}]{He2021}
Hellinger, P., Papini, E., Verdini, A., {et~al.} 2021, The Astrophysical
  Journal, 917, 101

\bibitem[{Hospodarsky {et~al.}(2017)Hospodarsky, Kurth, Bolton, Allegrini,
  Clark, Connerney, Ebert, Haggerty, Levin, McComas, {et~al.}}]{Ho2017}
Hospodarsky, G.~B., Kurth, W.~S., Bolton, S., {et~al.} 2017, Geophysical
  Research Letters, 44, 4506

\bibitem[{Huang {et~al.}(2020)Huang, Kasper, Vech, Klein, Stevens,
  Martinovi{\'c}, Alterman, {\v{D}}urovcov{\'a}, Paulson, Maruca,
  {et~al.}}]{Hu2020}
Huang, J., Kasper, J.~C., Vech, D., {et~al.} 2020, The Astrophysical Journal
  Supplement Series, 246, 70

\bibitem[{Huang {et~al.}(2017)Huang, Hadid, Sahraoui, Yuan, \& Deng}]{Hu2017}
Huang, S., Hadid, L., Sahraoui, F., Yuan, Z., \& Deng, X. 2017, The
  Astrophysical Journal Letters, 836, L10

\bibitem[{Huang {et~al.}(2014)Huang, Sahraoui, Deng, He, Yuan, Zhou, Pang, \&
  Fu}]{Huang2014ApJL}
Huang, S.~Y., Sahraoui, F., Deng, X.~H., {et~al.} 2014, The Astrophysical
  Journal Letters, 789, L28.
\newblock \url{http://stacks.iop.org/2041-8205/789/i=2/a=L28}

\bibitem[{Kivelson(1995)}]{K1995}
Kivelson, A. 1995, Introduction to space physics (Cambridge university press)

\bibitem[{MacBride {et~al.}(2008)MacBride, Smith, \& Forman}]{Mc2008}
MacBride, B.~T., Smith, C.~W., \& Forman, M.~A. 2008, Astrophys. J., 679, 1644

\bibitem[{Marino \& Sorriso-Valvo(2023)}]{Ma2023}
Marino, R., \& Sorriso-Valvo, L. 2023, Physics Reports, 1006, 1

\bibitem[{McComas {et~al.}(2017)McComas, Alexander, Allegrini, Bagenal, Beebe,
  Clark, Crary, Desai, De~Los~Santos, Demkee, {et~al.}}]{Mc2017}
McComas, D., Alexander, N., Allegrini, F., {et~al.} 2017, Space Science
  Reviews, 213, 547

\bibitem[{Mininni \& Pouquet(2009)}]{Mi2009}
Mininni, P.~D., \& Pouquet, A. 2009, Phys. Rev. E, 80, 025401

\bibitem[{Monin \& Yaglom(1975)}]{MY1975}
Monin, A.~S., \& Yaglom, A.~M. 1975, Statistical Fluid Mechanics: Mechanics of
  Turbulence, Vol.~2 (Cambridge, MA: MIT Press.)

\bibitem[{Politano \& Pouquet(1998a)}]{P1998a}
Politano, H., \& Pouquet, A. 1998a, Physical Review E, 57, R21

\bibitem[{Politano \& Pouquet(1998b)}]{P1998b}
---. 1998b, Geophysical Research Letters, 25, 273

\bibitem[{Qudsi {et~al.}(2020)Qudsi, Maruca, Matthaeus, Parashar,
  Bandyopadhyay, Chhiber, Chasapis, Goldstein, Bale, Bonnell, de~Wit, Goetz,
  Harvey, MacDowall, Malaspina, Pulupa, Kasper, Korreck, Case, Stevens,
  Whittlesey, Larson, Livi, Velli, \& Raouafi}]{Qudsi2020ApJS}
Qudsi, R.~A., Maruca, B.~A., Matthaeus, W.~H., {et~al.} 2020, The Astrophysical
  Journal Supplement Series, 246, 46, \dodoi{10.3847/1538-4365/ab5c19}

\bibitem[{Ranquist {et~al.}(2019)Ranquist, Bagenal, Wilson, Hospodarsky, Ebert,
  Allegrini, Valek, McComas, Connerney, Kurth, \& Bolton}]{Ranquist2019JGR}
Ranquist, D.~A., Bagenal, F., Wilson, R.~J., {et~al.} 2019, Journal of
  Geophysical Research: Space Physics, 124, 9106, \dodoi{10.1029/2019JA027382}

\bibitem[{Romanelli {et~al.}(2022)Romanelli, Andres, \& DiBraccio}]{R2022}
Romanelli, N., Andres, N., \& DiBraccio, G. 2022, accepted in ApJ. ArXiv
  preprint arXiv:2201.11772

\bibitem[{Russell(1993)}]{R1993}
Russell, C. 1993, Reports on Progress in Physics, 56, 687

\bibitem[{Simon \& Sahraoui(2021)}]{S2021}
Simon, P., \& Sahraoui, F. 2021, The Astrophysical Journal, 916, 49

\bibitem[{Sorriso-Valvo {et~al.}(2007)Sorriso-Valvo, Marino, Carbone, Noullez,
  Lepreti, Veltri, Bruno, Bavassano, \& Pietropaolo}]{SV2007}
Sorriso-Valvo, L., Marino, R., Carbone, V., {et~al.} 2007, Physical review
  letters, 99, 115001

\bibitem[{Taylor(1938)}]{Taylor1938PRSLA}
Taylor, G.~I. 1938, Proceedings of the Royal Society of London Series A, 164,
  476, \dodoi{10.1098/rspa.1938.0032}

\bibitem[{von K\'arm\'an \& Howarth(1938)}]{vkh1938}
von K\'arm\'an, T., \& Howarth, L. 1938, Proceedings of the Royal Society of
  London A: Mathematical, Physical and Engineering Sciences, 164, 192,
  \dodoi{10.1098/rspa.1938.0013}

\end{thebibliography}

\end{document}